\documentclass[a4paper,USenglish]{lipics-v2016}

\usepackage{url}
\usepackage{amssymb}
\usepackage{graphicx}
\usepackage{amsmath}
\usepackage{amsfonts}
\usepackage{algorithm}
\usepackage{algpseudocode}
\usepackage{color}

\title{Efficient and Effective Quantum Compiling for Entanglement-based Machine Learning on IBM Q Devices}
\titlerunning{Quantum Compiling for Entanglement-based Machine Learning on IBM Q Devices}

\author{Davide Ferrari, Michele Amoretti}
\affil{Department of Engineering and Architecture - University of Parma, Italy} 
\affil{Quantum Information Science - University of Parma, Italy}
\affil{Contact: michele.amoretti@unipr.it}
\authorrunning{D. Ferrari, M. Amoretti} 

\keywords{Quantum Compiling; IBM Q; Parity Learning}

\begin{document}

\maketitle

\begin{abstract}
Quantum compiling means fast, device-aware implementation of quantum algorithms (i.e., quantum circuits, in the quantum circuit model of computation).
In this paper, we present a strategy for compiling IBM Q -aware, low-depth quantum circuits that generate Greenberger-Horne-Zeilinger (GHZ) entangled states. The resulting compiler can replace the QISKit compiler for the specific purpose of obtaining improved GHZ circuits. It is well known that GHZ states have several practical applications, including quantum machine learning. We illustrate our experience in implementing and querying a uniform quantum example oracle based on the GHZ circuit, for solving the classically hard problem of learning parity with noise.
\end{abstract}

\section{Introduction}
\label{sec:intro}

Since 2016, IBM offers hands-on, cloud-based access to its experimental quantum computing platform, denoted as IBM Q Experience \cite{IBMQE}. Such a platform comprises a 5 qubit device, denoted as QX4, and a 16 qubit device, named QX5. All devices are based on transmon qubits \cite{Koch2007}, i.e., superconducting qubits that are insensitive to charge noise and have $\sim 100\mu$sec decoherence time. 

IBM Q Experience is an effective experimental platform for testing quantum algorithms. A Web-based visual tool provides a convenient way to compose quantum circuits for the 5 qubit device and run either simulated or real experiments. Alternatively, circuits can be designed by means of the QASM language and experiments can be defined and executed by means of the QISKit Python SDK \cite{QISKit}. Actually, IBM QX5 can be accessed in this way only. QISKit provides a quite large set of quantum gates for the user's convenience. However, only three gates are physically implemented by IBM Q, while all other gates are obtained as combinations of the former ones \cite{Coles2018}.

Quantum compiling means fast, device-aware implementation of quantum algorithms. With respect to IBM Q devices, compiling is not just a matter of translating virtual gates into physical ones. There are also device-specific physical constraints that prevent from placing CNOT gates wherever they are needed. QISKit comes with a default compiler, based on a randomized algorithm\cite{Bishop2017}. Another framework denoted as ProjectQ \cite{Steiger2018} has its own compiler \cite{Haner2018} that is capable of targeting various types of hardware, including IBM Q. Both QISKit and ProjectQ compilers are general-purpose, but the quantum circuits they build are not always optimal, especially when many qubits are involved. Namely, the resulting circuits may be characterized by too great \textit{depth}, which is defined as the number of layers of gates acting simultaneously on disjoint sets of qubits \cite{Motzoi2017}. In other words, the depth of a quantum circuit is the longest of the paths between each qubit and its measurement gate. Low-depth circuits are preferable, as they provide faster computations, meaning that results are less affected by quantum decoherence.

In this paper, we present a strategy for compiling IBM Q -aware, low-depth quantum circuits that generate  Greenberger-Horne-Zeilinger (GHZ) entangled states \cite{GHZ1989} (shortly, GHZ circuits).  
The Python implementation of our compiler can replace the QISKit compiler for the specific purpose of obtaining improved GHZ circuits. It is well known that GHZ states have several practical applications, including quantum machine learning \cite{Cai2015,Cross2015,Riste2017}.

Machine learning techniques are powerful tools for finding patterns in data. The field of quantum machine learning looks for faster machine learning algorithms, based on quantum computing principles \cite{Biamonte2017}. Its cornerstones are the HHL algorithm \cite{Harrow2009} for (approximately) solving systems of linear equations and the ``learning from quantum examples'' approach \cite{Bshouty1999,Atici2005,Zhang2010,Arunachalam2017}, each example being a coherent quantum state.

In this paper we focus on parity learning, which is a machine learning problem that is easy to solve, with a classical computer, only in the noiseless case. The noisy version of the problem, instead, is conjectured to be classically hard. For this reason, it has recently found many applications in cryptography \cite{Regev2005,Pietrzak2012}. However, Cross \textit{et al.} \cite{Cross2015} proved that learning parity with noise can be performed with superpolynomial quantum computational speedup. The experimental demonstration on IBM QX2 was recently presented by Rist\`e \textit{et al.}\cite{Riste2017} In this work, we illustrate similar experiments we have performed on IBM QX5. GHZ circuits play a major role in these quantum computations.

The paper is organized as follow. 
In Section \ref{sec:qiskit-compiler}, we analyze the QISKit compiler.
In Section \ref{sec:proposed-strategy}, we illustrate our strategy for compiling GHZ circuits. 
In Section \ref{sec:eval-GHZ}, we compare the GHZ circuits produced by the QISKit compiler with those built by the proposed compiler.
In Section \ref{sec:learning}, we summarize the principles of quantum learning robust to noise. 
In Section \ref{sec:exp-learn}, we illustrate the experimental demonstration of parity learning by querying a uniform quantum example oracle. Finally, in Section \ref{sec:conclusion}, we conclude the paper with a discussion of open issues and future work.

\section{QISKit compiler}
\label{sec:qiskit-compiler}
\begin{figure}[!ht]
\centering
\includegraphics[width=3cm]{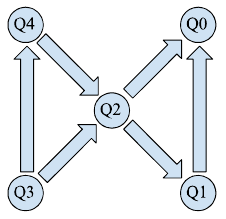}
\caption{IBM QX4 coupling map \cite{IBMQX4}, where an arrow from qubit A to qubit B means that A can act as the control qubit for a CNOT gate with target qubit B.}
\label{fig:ibmQX4cm}
\end{figure}

\begin{figure}[!ht]
\centering
\includegraphics[width=12cm]{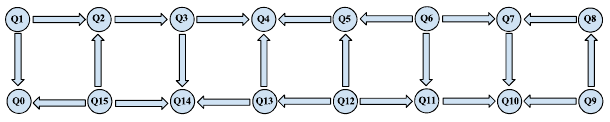}
\caption{IBM QX5 coupling map \cite{IBMQX5}.}
\label{fig:ibmQX5cm}
\end{figure}

QISKit\footnote{We refer to QISKit version 0.5.7 released the 20th of July 2018.} provides a \textsf{compile()} function to translate any ideal quantum circuit designed by the programmer to an equivalent quantum circuit that is suitable for the chosen IBM Q device. Each device is characterized by a coupling map, i.e., a directed graph representing superconducting bus connections between qubit pairs, which can be seen as the possibility to place two-qubits gates, like CNOT, between those qubits.
The coupling maps of IBM QX4 and QX5 are illustrated in Figures \ref{fig:ibmQX4cm} and \ref{fig:ibmQX5cm}, respectively.

QISKit's function \textsf{compile()} includes:
\begin{enumerate}
\item function \textsf{pick\_best\_layout()} --- finds the $n$ best physical qubits, taking into account the coupling map of the chosen backend (the best physical qubits are those that are most connected); associates the $n$ program-defined qubits to the $n$ best physical qubits obtained at the previous step;
\item function \textsf{transpile()} --- returns the compiled circuit, using:
\begin{itemize}
\item function \textsf{swap\_mapper()} --- returns a circuit that is equivalent to the program-defined one and respects the coupling map (modulo the fact that some CNOT gates have to be inverted);
\item function \textsf{direction\_mapper()} --- inverts the CNOT gates when necessary.
\end{itemize}
\end{enumerate}

To find the $n$ best physical qubits, function \textsf{pick\_best\_layout()} applies breadth-first search (BFS) to each vertex of the coupling map. The running time of BFS is $O(V+E)$, where $V$ is the number of vertices and $E$ is the number of edges of the considered coupling map. 
Usually, coupling maps have $E \sim m$ and $V \sim m$, where $m$ is the number of available qubits in the considered device (e.g., $m=16$ in QX5). Thus, the running time is $O(m^2)$. The association of the $n$ program-defined qubits to the $n$ best physical qubits is a simple loop, with running time $O(n)$. Finally, we can state that the running time of \textsf{pick\_best\_layout()} is $O(m^2)$.

Function \textsf{swap\_mapper()} performs \textsf{layer\_permutation()} on each layer of the circuit. 
The goal of \textsf{layer\_permutation()} is to swap qubits such that qubits in the same two-qubit gates are adjacent. First, it is checked whether 2-qubit gates can be directly applied according to current layout, in the considered layer. The running time of this operation is $O(n)$, as the maximum number of gates to check is $n/2$, for each layer. If the check is not successful, a randomized algorithm \cite{Bishop2017} is applied to find a correct permutation of the qubits, to satisfy the adjacency constraint. The randomized algorithm starts with the creation of a matrix of randomized distances between nodes of the coupling map. Creating such a matrix requires $O(n^2)$ steps. Then, there are three nested loops: the first one over $L \in \{1,..,2n+1\}$, being $L$ the number of layers of the swap circuit; the second one over the qubit set involved in the considered layer, whose size is $O(n)$; the third one is over the set of edges in the coupling map (recall that $E \sim m$). The randomized algorithm is repeated $T$ times ($T=20$ by default), in order to minimize an objective function defined as the sum of the randomized distances between the qubits involved in the trial layout.
Thus, the running time of the randomized algorithm is $O(Tn^2m)$. 
Supposing there are $L$ layers to be checked, the running time of \textsf{swap\_mapper()} is $O(LTn^2m)$.
Being $L \sim n$, the running time of \textsf{swap\_mapper()} is $O(Tn^3m)$.

Considering $\sim n$ layers and at most $n/2$ CNOT gates for each layer, the running time of \textsf{direction\_mapper()} turns out to be $O(n^2)$.

To summarize, the running time of the QISKit compiler is $O(Tn^3m)$. It is a universal compiler, coping with any quantum circuit, but it does not guarantee to find an optimal solution (because of randomization), and for large circuits it may be quite inefficient.

\section{Proposed compiling strategy for GHZ circuits}
\label{sec:proposed-strategy}

\begin{figure}[!ht]
\centering
\includegraphics[width=8cm]{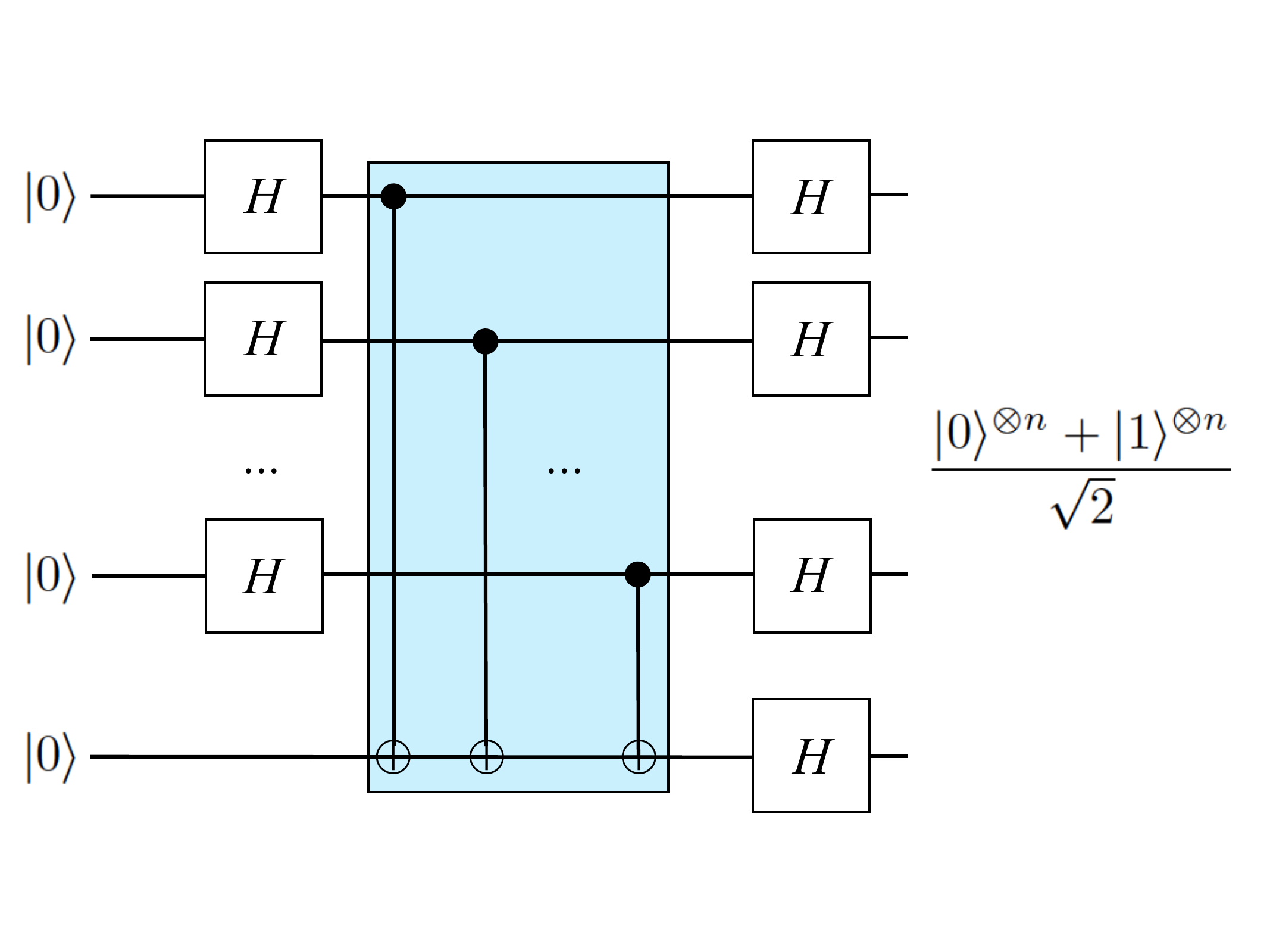}
\caption{Ideal GHZ circuit \cite{Deffner2017}.}
\label{fig:GHZ}
\end{figure}

In the Greenberger-Horne-Zeilinger (GHZ) entangled state \cite{GHZ1989}, either all of the $n$ qubits are in the zero state or all of them are in the one state:
\begin{equation}
\frac{|0\rangle^{\otimes n} + |1\rangle^{\otimes n}}{\sqrt{2}}
\end{equation}
Such a state plays a prominent role in several quantum algorithms.
To generate the GHZ state, one may use the quantum circuit illustrated in Figure \ref{fig:GHZ} \cite{Deffner2017}, which does require $H$ and $CNOT$ gates only. 

The coupling maps of IBM Q devices prevent from placing all CNOT gates like in the ideal GHZ circuit, i.e., making $n-1$ qubits control the $n$-th qubit. Fortunately, it is possible to turn the ideal GHZ circuit to an equivalent one characterized by a unique CNOT gate sequence, where the $(j+1)$-th qubit controls the $j$-th qubit \cite{Ozaeta2017}. Taking advantage of this property, our deterministic compiling strategy builds a GHZ circuit by placing CNOT gates (and inverse-CNOT gates) according to a low-depth spanning tree built over the coupling map. The resulting GHZ circuit is characterized by low depth. 

The proposed approach is highly modular. The first two steps are performed once for each IBM Q device, which is characterized by $m$ qubits (e.g., $m=16$ in QX5). The subsequent steps, instead, are specific for building GHZ circuits with $n \leq m$ qubits.

More precisely, the first step consists in finding the physical qubit that can be reached by as many other qubits as possible through directed paths in the coupling map. The running time of this search is $O(m^2)$. Starting from the resulting qubit, the second step consists in creating a low-depth spanning tree connecting all the qubits of the device. The running time of this process is $O(m)$. The resulting spanning tree can be saved to a file for being reused any time.
The GHZ-specific steps consist in 
\begin{itemize}
\item placing CNOT gates (and inverse-CNOT gates), leveraging the spanning tree (running time $O(n)$);  
\item removing $HH = I$ gate sequences, if any (running time $O(Dn)$, where $D$ is the depth of the circuit obtained so far).
\end{itemize}

The Python implementation of the proposed GHZ circuit compiler is available as open source on GitHub.\footnote{\url{https://github.com/qis-unipr/QuantumComputing/}} In the following, we describe its main functions with more details.

Every node $x$ of the coupling map is assigned a rank $R[x]$, defined as the number of nodes that can reach $x$ along the directed edges of the coupling map. The node with the highest rank is then selected as the starting point for building the circuit.

Recursive function \textsf{explore()} (Algorithm \ref{alg:explore}) starts from a source node $s$ and explores the paths that go from there. Being the coupling map a directed graph, each node has in-neighbors (predecessors) and out-neighbors (successors). The node being visited is denoted as $v$ and its successors are denoted as $\mathcal{S}_v$. For each node $x \in \mathcal{S}_v$, if it does not belong to the set $\mathcal{V}_s$ of visited nodes associated to the source node $s$, $x$ is put into $\mathcal{V}_s$ and increment its global rank $R[x]$ by one. In this way, the search for nodes that can be reached from $s$ is exhaustive, but no one node is explored more than once.

\begin{algorithm}
\footnotesize
\caption{\textsf{explore($s$,$v$,$R$)}}
\label{alg:explore}
\begin{algorithmic}
\ForAll{$x \in \mathcal{S}_v$}
\If{$x \not\in \mathcal{V}_s$}
\State put $x$ into $\mathcal{V}_s$
\State $R[x] \gets R[x]+1$
\State \textsf{explore($s$,$x$,$R$)}
\EndIf
\EndFor
\end{algorithmic}
\end{algorithm}

As soon as the qubit with the highest rank has been found, function \textsf{spanning\_tree()} (Algorithm \ref{alg:spanningtree}) is executed, in order to obtain a spanning tree connecting all the $m$ available qubits. There, $s$ denotes the source node (the one with highest rank $R$), $\mathcal{P}$ is the set of predecessors of a given node, $\mathcal{C}$ is the set of nodes to be connected, and $\mathcal{T}$ is the set of node pairs corresponding to the desired spanning tree.  

\begin{algorithm}
\footnotesize
\caption{\textsf{spanning\_tree($s$, $\mathcal{R}$)}}
\label{alg:spanningtree}
\begin{algorithmic}
\State $\mathcal{C} \gets \emptyset$
\State $\mathcal{T} \gets \emptyset$
\State put $s$ into $\mathcal{C}$
\State $\mathcal{R} \gets \mathcal{R} / s$
\ForAll{$v \in \mathcal{C}$}
\ForAll{$x \in \mathcal{P}_v$}
\If{$x \not\in \mathcal{T}$}
\State put $(x,v)$ into $\mathcal{T}$
\State $\mathcal{R} \gets \mathcal{R} / x$
\If{$x \not\in \mathcal{C}$}
\State put $x$ into $\mathcal{C}$
\EndIf
\If{$|\mathcal{R}| = 0$}
\State return $\mathcal{T}$
\EndIf
\EndIf
\EndFor
\If{$v$ is last}
\ForAll{$r \in \mathcal{R}$}
\ForAll{$y \in \mathcal{P}_r$}
\If{$y \in \mathcal{C}$}
\State put $r$ into $\mathcal{C}$
\State put $(r,y)$ into $\mathcal{T}$
\State $\mathcal{R} \gets \mathcal{R} / r$
\If{$|\mathcal{R}| = 0$}
\State return $\mathcal{T}$
\EndIf
\State break
\EndIf
\EndFor
\If{$v$ is not last}
\State break
\EndIf
\EndFor
\EndIf
\EndFor
\end{algorithmic}
\end{algorithm}

\begin{figure}[!ht]
\centering
\includegraphics[width=6cm]{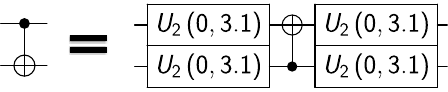}
\caption{Inverse-CNOT circuit. $U_2(0,3.1)$ stands for $H$.}
\label{fig:inverseCNOT}
\end{figure}

Function \textsf{place\_cnot()} (Algorithm \ref{alg:placecnot}) walks the aforementioned path and uses the \textsf{cnot()} function (Algorithm \ref{alg:cnot}) to put across each node pair either a CNOT or an inverse-CNOT gate (illustrated in Figure \ref{fig:inverseCNOT}), depending on the direction of the link dictated by the coupling map. In \textsf{cnot()}, $\mathcal{S}_c$ is the set of successors of node $c$.

\begin{algorithm}
\footnotesize
\caption{\textsf{place\_cnot($\mathcal{T}$,$n$)}}
\label{alg:placecnot}
\begin{algorithmic}
\State $\tau \gets n-1$
\ForAll{$(x_1,x_2) \in \mathcal{T}$}
\State \textsf{cnot($x_1$,$x_2$)}
\State $\tau \gets \tau - 1$
\If{$\tau = 0$}
\State break
\EndIf
\EndFor
\end{algorithmic}
\end{algorithm}

\begin{algorithm}
\footnotesize
\caption{\textsf{cnot($c$,$t$)}}
\label{alg:cnot}
\begin{algorithmic}
\If{$t \in \mathcal{S}_c$}
\State apply CNOT with $c$ as control and $t$ as target
\Else
\State apply H gates to $c$ and $t$ 
\State apply CNOT with $t$ as control and $c$ as target
\State apply H gates to $c$ and $t$ 
\EndIf
\end{algorithmic}
\end{algorithm}

\begin{figure}[!ht]
\centering
\includegraphics[width=3cm]{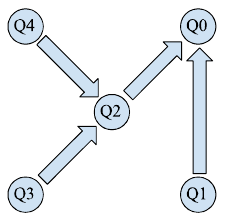}
\includegraphics[width=12cm]{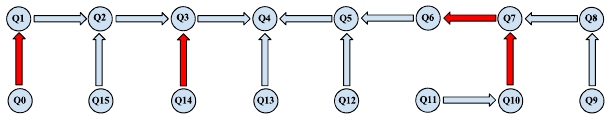}
\caption{Couplings between qubits in the resulting circuits, for QX4 and QX5 respectively. Red arrows correspond to the presence of inverse-CNOT gates.}
\label{fig:adjustedCMs}
\end{figure}

In Figure \ref{fig:adjustedCMs}, the couplings between qubits in the resulting circuits are illustrated, for QX4 and QX5 respectively. With our approach, the number of gates is kept to a minimum, which is good as every gate added to the circuit brings a certain amount of error with it.

\section{Experimental evaluation of compiled GHZ circuits}
\label{sec:eval-GHZ}

To compare the proposed compiler with the QISKit compiler in generating GHZ circuits, we use two metrics. The first one is circuit depth $d$, i.e., the longest of the paths between each qubit and its measurement gate. The second one is classical fidelity coefficient $B$ (Bhattacharyya coefficient), evaluating the classical output generated by the compiled circuit with respect to the theoretically expected one.

\begin{figure}[!ht]
\tiny QISKit compiler\\
\includegraphics[width=12cm]{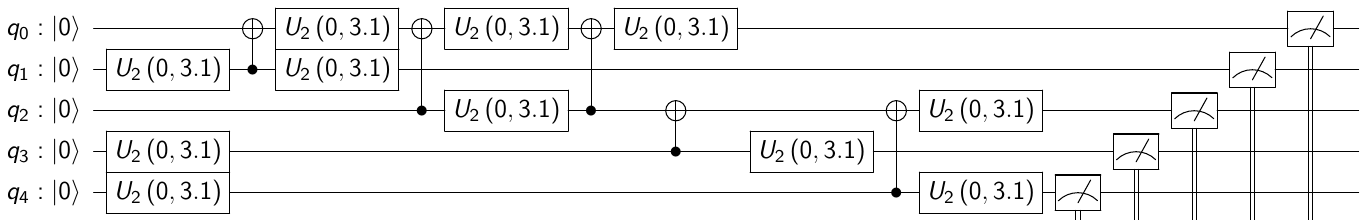}\\
\tiny Proposed compiler\\
\includegraphics[width=8cm]{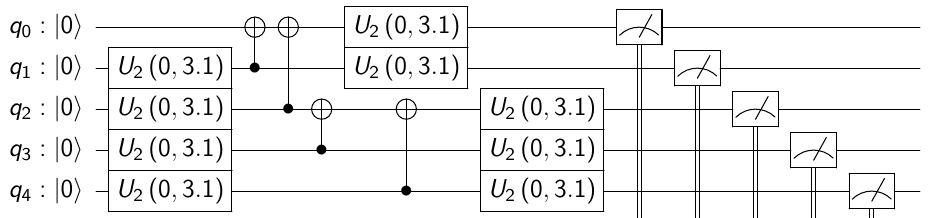}
\caption{GHZ circuits with $n=5$ qubits produced by QISKit compiler and by the proposed compiler, for the QX4 device.}
\label{fig:GHZcircuitsQX4}
\end{figure}

\begin{figure}[!ht]
\tiny QISKit compiler\\
\includegraphics[width=12.5cm]{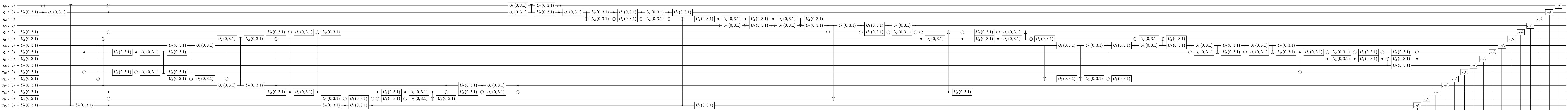}
\tiny Proposed compiler\\
\includegraphics[width=3.3cm]{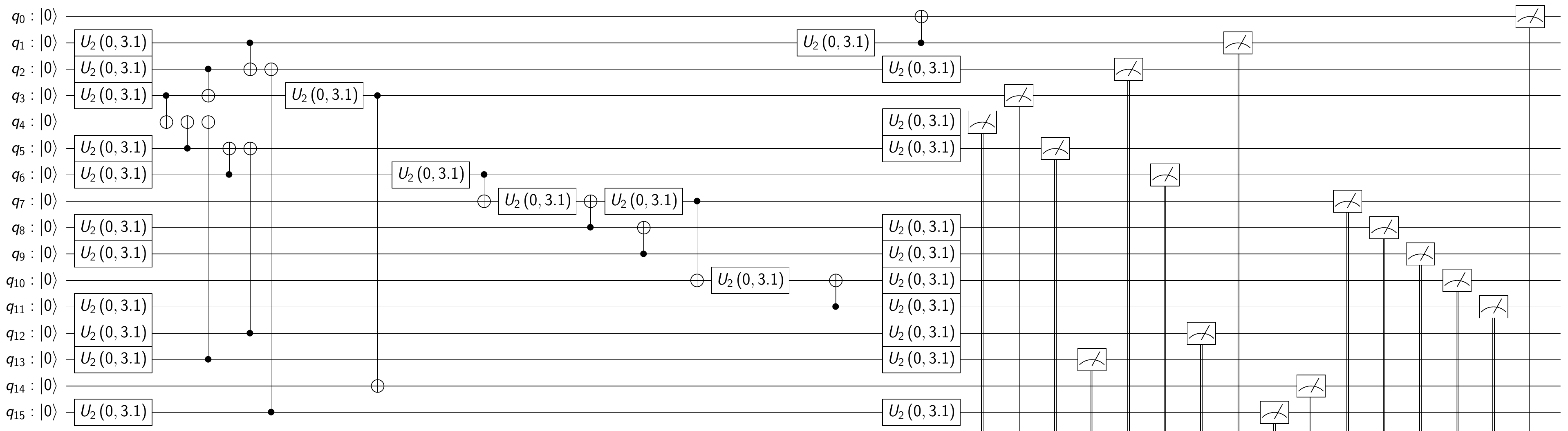}
\caption{GHZ circuits with $n=16$ qubits produced by QISKit compiler and by the proposed compiler, for the QX5 device.}
\label{fig:GHZcircuitsQX5}
\end{figure}

In Figures \ref{fig:GHZcircuitsQX4} and \ref{fig:GHZcircuitsQX5}, some GHZ circuits generated by the QISKit compiler are compared with those built by the proposed compiler. Regarding the QISKit compiler, we show the best circuits among those we obtained with repeated trials, until we achieved a sufficiently narrow $I_{95}$ confidence interval for $d$. Conversely, the proposed deterministic compiler required only one execution for each $n$ value.

\begin{figure}[!ht]
\centering
\includegraphics[width=6cm]{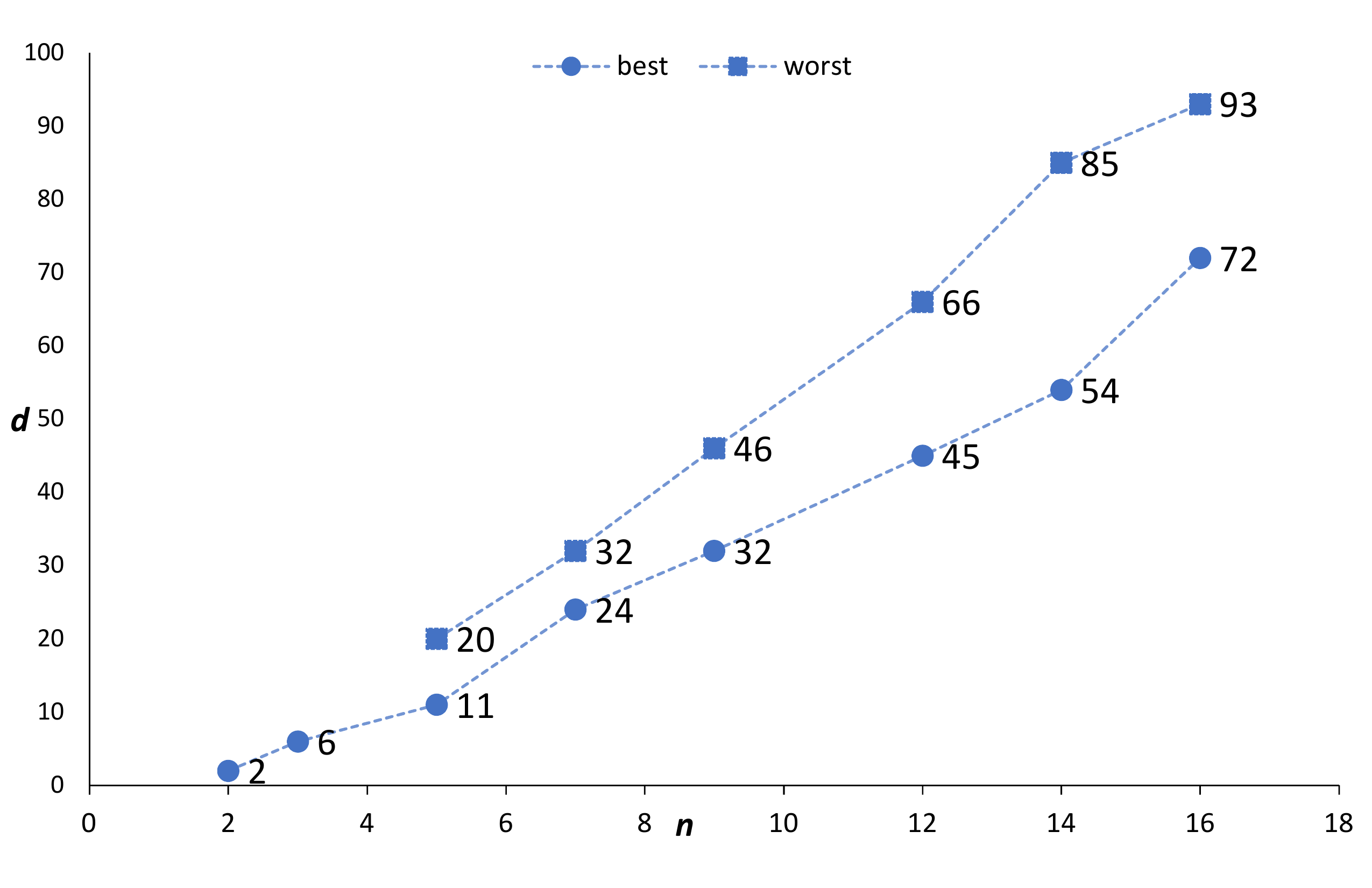}
\includegraphics[width=6cm]{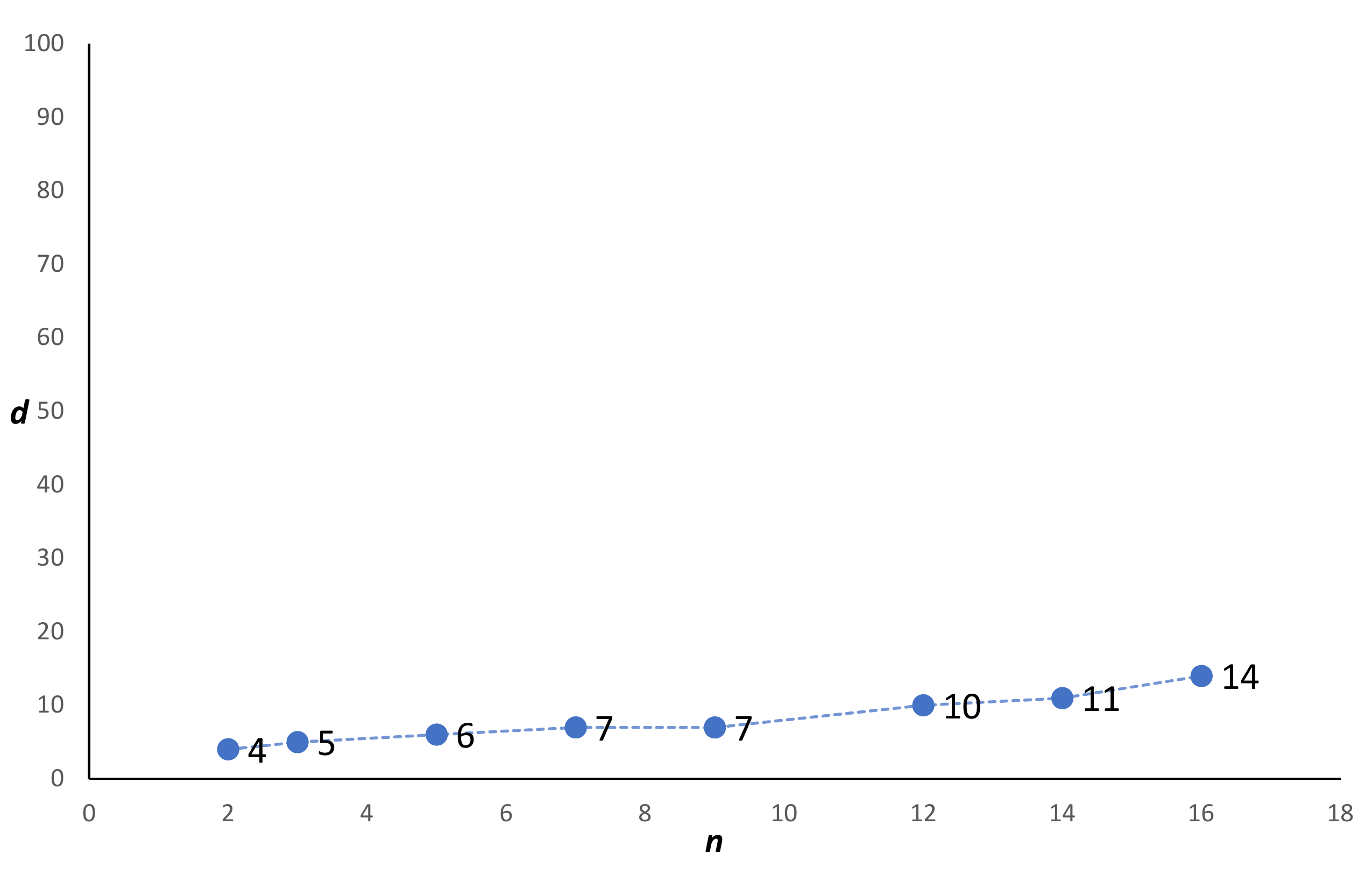}
\caption{Depth of best and worst GHZ circuits built by the QISKit compiler (left) and depth of GHZ circuits produced by the proposed compiler (right), for increasing $n$ values on QX5.}
\label{fig:d-vs-n}
\end{figure}

In Figure \ref{fig:d-vs-n}, the depth of the best and worst circuits produced by the QISKit compiler is compared to the fixed depth of the circuits built by the proposed compiler, versus $n$. It is evident that the GHZ circuits built by the proposed compiler have $d \sim n$, while those produced by the QISKit compiler have $d \sim 5n$.

\begin{figure}[!ht]
\centering
\includegraphics[width=6cm]{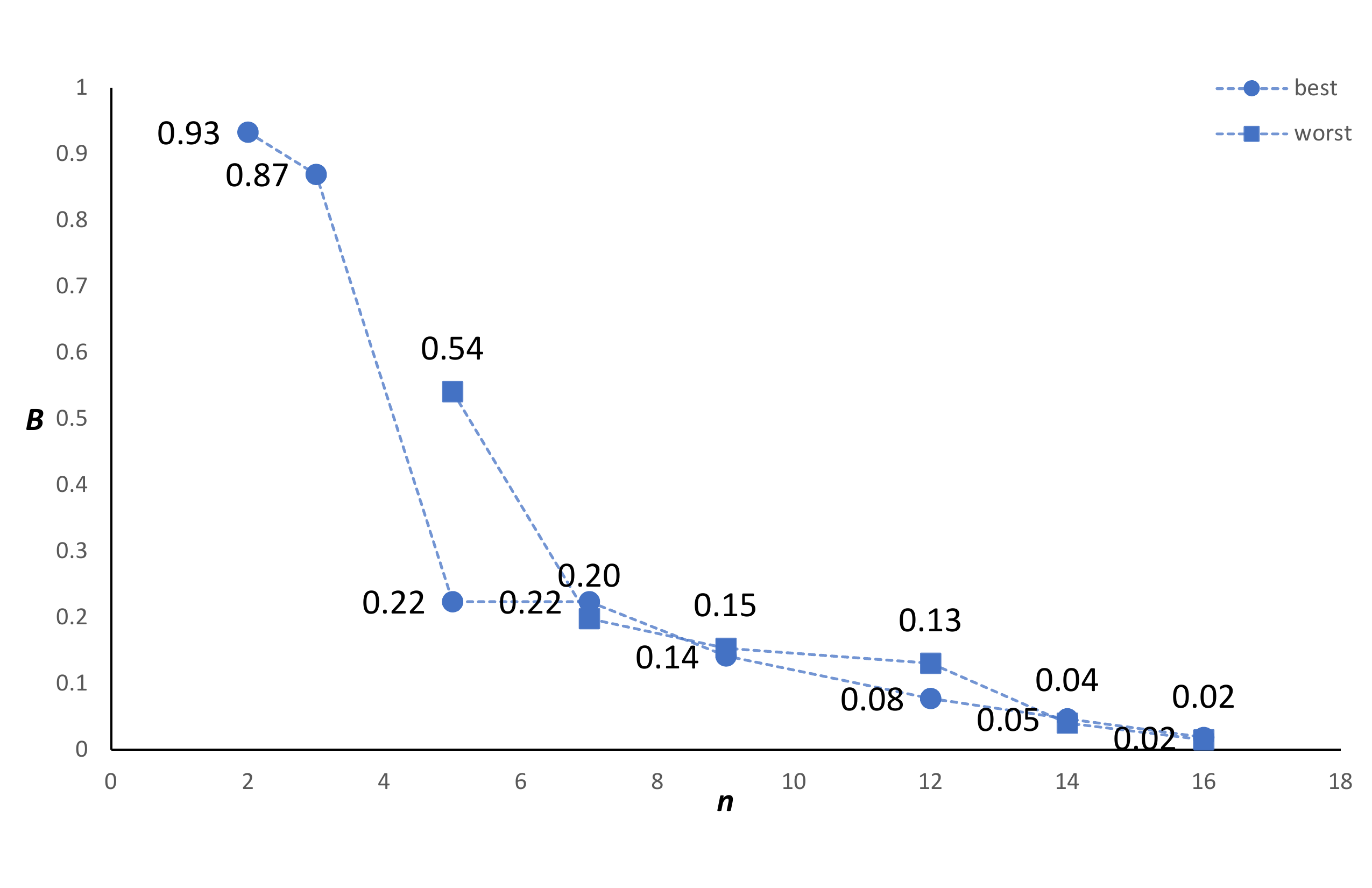}
\includegraphics[width=6cm]{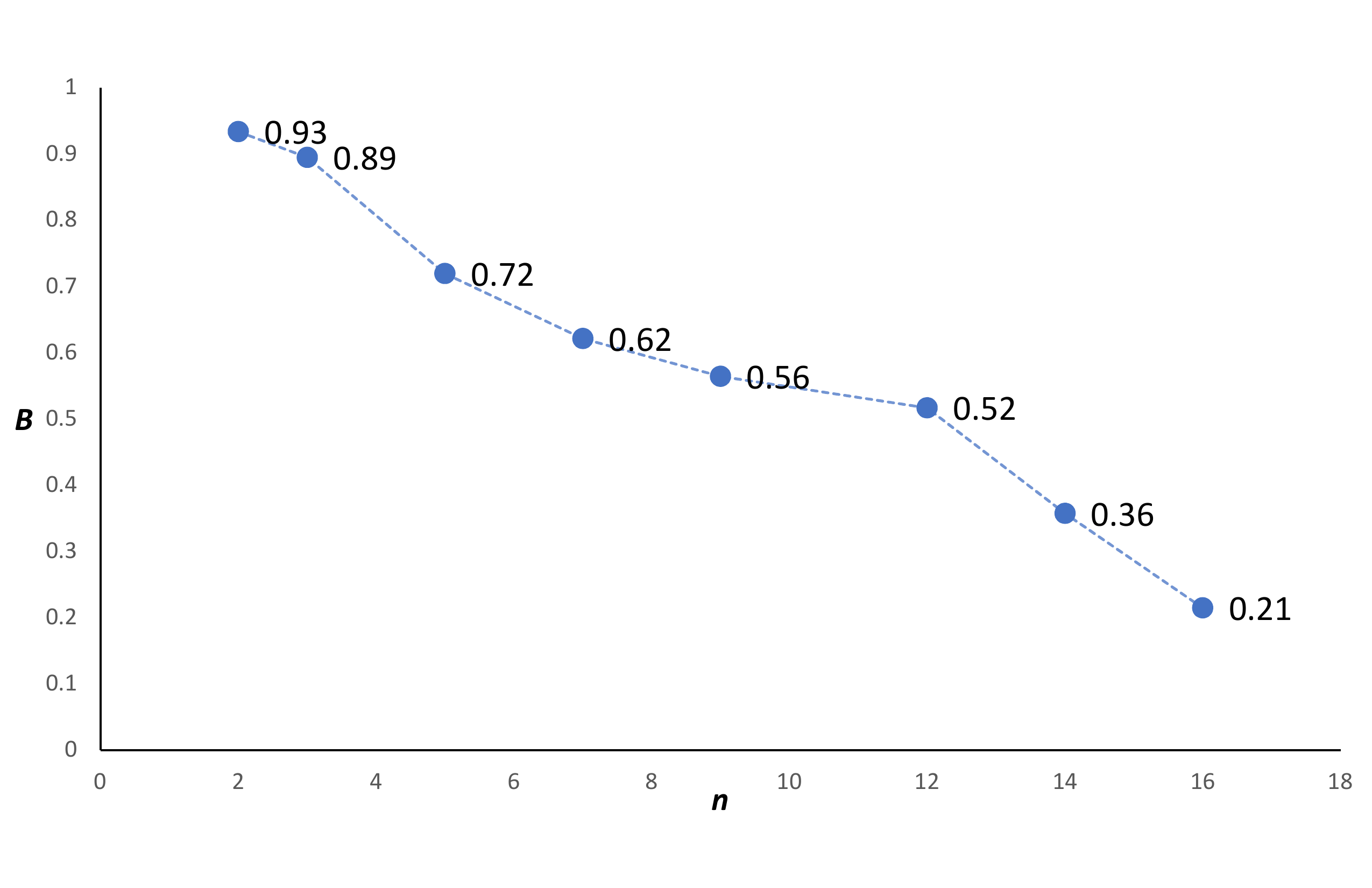}
\caption{Classical fidelity of GHZ states generated by the circuits built by the QISKit compiler (left) and by the proposed compiler (right), for increasing $n$ values on QX5.}
\label{fig:fidelity}
\end{figure}

In Figure \ref{fig:fidelity}, it is shown that the GHZ states generated by the circuits built by the proposed compiler are characterized by higher classical fidelity than those deriving from the QISKit compiler (considering 10 executions, each one consisting of 8192 shots for each $n$ value, on QX5).

\section{Quantum learning robust to noise}
\label{sec:learning}

Binary classification is a particular learning task aiming to identify an unknown mapping $f:\{0,1\}^n \rightarrow \{0,1\}$.
The problem can be solved by means of an oracle, i.e., an efficient, simple to build circuit that accepts an input $x \in \{0,1\}^n$ and returns $f(x)$.
With a quantum oracle, a superposition of all possible input states $|x\rangle$ can be used to obtain a superposition of all respective $|f(x)\rangle$ states. The corresponding classical oracle is obtained by measuring the output state. The classical learner makes repeated requests and measurements, and uses the obtained output bits, together with classical computation, to learn $f$.
However, in some cases, clever use of the quantum oracle in fully quantum settings may result in a highly reduced query count with respect to classical algorithms \cite{Cross2015,Riste2017}.

\begin{figure}[!ht]
\centering
\includegraphics[width=7cm]{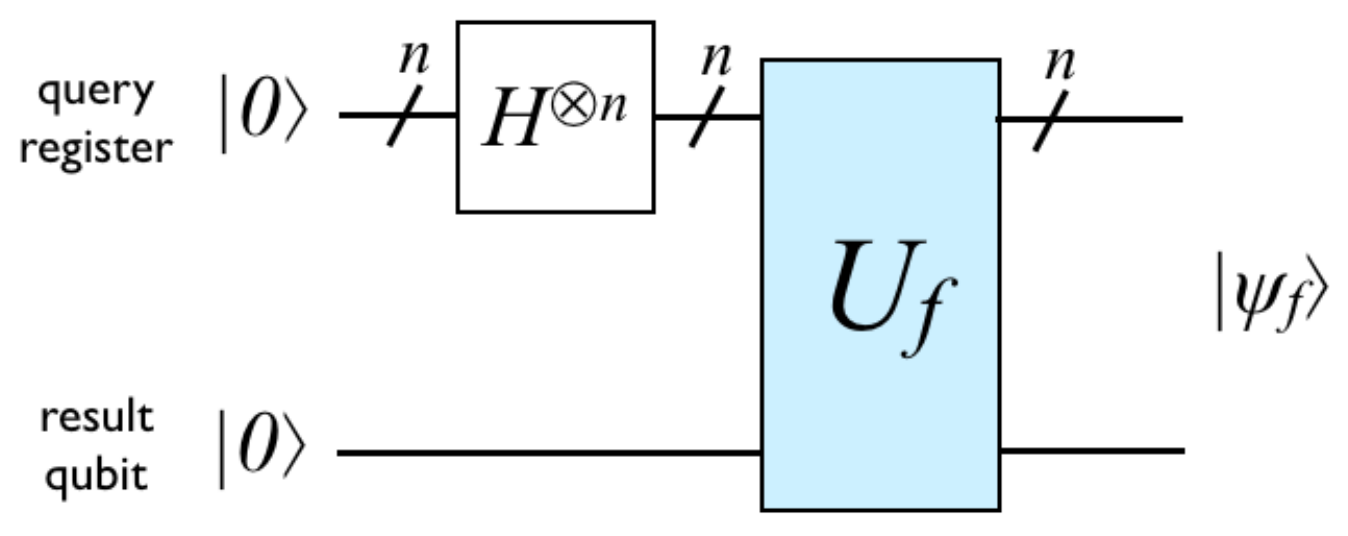}
\caption{Uniform quantum example oracle for $f$. $H$ denotes the Hadamard gate.}
\label{fig:oracle}
\end{figure}

We start by reviewing relevant definitions.
A uniform quantum example oracle\cite{Bshouty1999} for the Boolean function $f:\{0,1\}^n \rightarrow \{0,1\}$ is a unitary transformation that outputs the quantum state
\begin{equation}
|\psi_f\rangle = \frac{1}{2^{n/2}} \sum_{x \in \{0,1\}^n} |x, f(x)\rangle.
\end{equation}
Each learner's request to this oracle for a quantum state has unit cost. The \textit{query register} is the set of qubits that contain $x$, while the \textit{result qubit} is the auxiliary qubit that contains $f(x)$ (Figure \ref{fig:oracle}).

In learning theory \cite{Valiant1984,Natarajan1991}, function $f$ defined above represents a \textit{concept}. A collection of concepts is a \textit{concept class}. Given a \textit{target concept} $f$, a typical goal is constructing an \textit{$\epsilon$-approximation of $f$}, i.e., a function $h:\{0,1\}^n \rightarrow \{0,1\}$ that agrees with $f$ on at least a $1-\epsilon$ fraction of the inputs:
\begin{equation}
P[h(x)=f(x)] \geq 1 - \epsilon
\end{equation}

We now focus on the class of parity functions
\begin{equation}
f_a(x) = \langle a,x \rangle = \sum_{j=1}^n a_j x_j \mod 2
\end{equation}
where $a \in \{0,1\}^n$ and $a_j$ (respectively, $x_j$) denotes the $j$th bit of $a$ (respectively, $x$). 
The purpose of the learner is to find $a$ exactly \cite{Fischer1992,Hembold1992}. 

Firstly, let us consider the noiseless case.
The classical oracle must be queried $N=n$ times to learn $a$ exactly with a constant probability of success. In the quantum setting, $a$ can be learned with constant probability from a single query. It is sufficient to apply a Hadamard gate $H$ to each of the $n+1$ qubits, downstream of the oracle, thus obtaining the following output state:
\begin{equation}
\label{eq:output}
\frac{1}{\sqrt{2}}(|0^n,0\rangle + |a,1\rangle).
\end{equation}
Thus, measurement reveals $a$ whenever the result is $1$, with probability $1/2$. 

Learning parity with noise 	\cite{Kearns1998}, in the classical setting, is an average case version of the NP-hard problem of decoding a linear code \cite{Berlekamp1978}, which is also hard to approximate \cite{Hastad2001}. The conjectured hardness of the problem has found many applications in cryptography \cite{Regev2005,Pietrzak2012}. However, Cross \textit{et al.} \cite{Cross2015} have recently proved that learning $a$ in the quantum setting requires a number of queries $N$ that scales as $O(\log n)$. This result has been experimentally demonstrated on IBM QX2 by Rist\`e \textit{et al.} \cite{Riste2017}

\section{Experimental demonstration of parity learning}
\label{sec:exp-learn}

\begin{figure}[!ht]
\centering
\includegraphics[width=12cm]{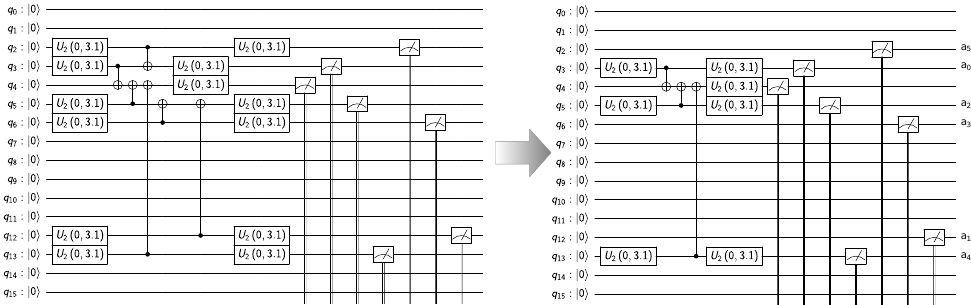}
\caption{Building the quantum oracle that encodes $a=101010$.}
\label{fig:oracle-example}
\end{figure}

The quantum parity oracle encoding $a = 11..11$ plus $H$ gates before measurement for quantum learning corresponds to the GHZ circuit illustrated in Figure \ref{fig:GHZ}, the target qubit playing the role of result qubit. With that circuit, to implement any $a$ sequence it would be sufficient to remove the CNOT gates corresponding to the qubits we want to bring to state $|0\rangle$. However, in the GHZ circuit produced by our compiler, CNOT gates form chains corresponding to branches of the spanning tree. Thus, removing an arbitrary CNOT would bring all qubits involved in the subsequent CNOT gates to state $|0\rangle$. What we do instead is to selectively remove the CNOT gates from the end of the branches of the spanning tree and to change the order in which the qubits are read, so that the zeroes are in the correct position. In Figure \ref{fig:oracle-example}, we show the encoding of $a=101010$ on QX5. Starting from the quantum circuit that encodes $a=111111$, we remove the CNOT gates that are controlled by qubits $q_2q_6q_{12}$, as they are the most peripheral. Then, we measure and read the results in the following qubit order: $q_3q_{12}q_5q_6q_{13}q_2$.

\begin{figure}[!ht]
$a=00..00$\\ 
\includegraphics[width=6cm]{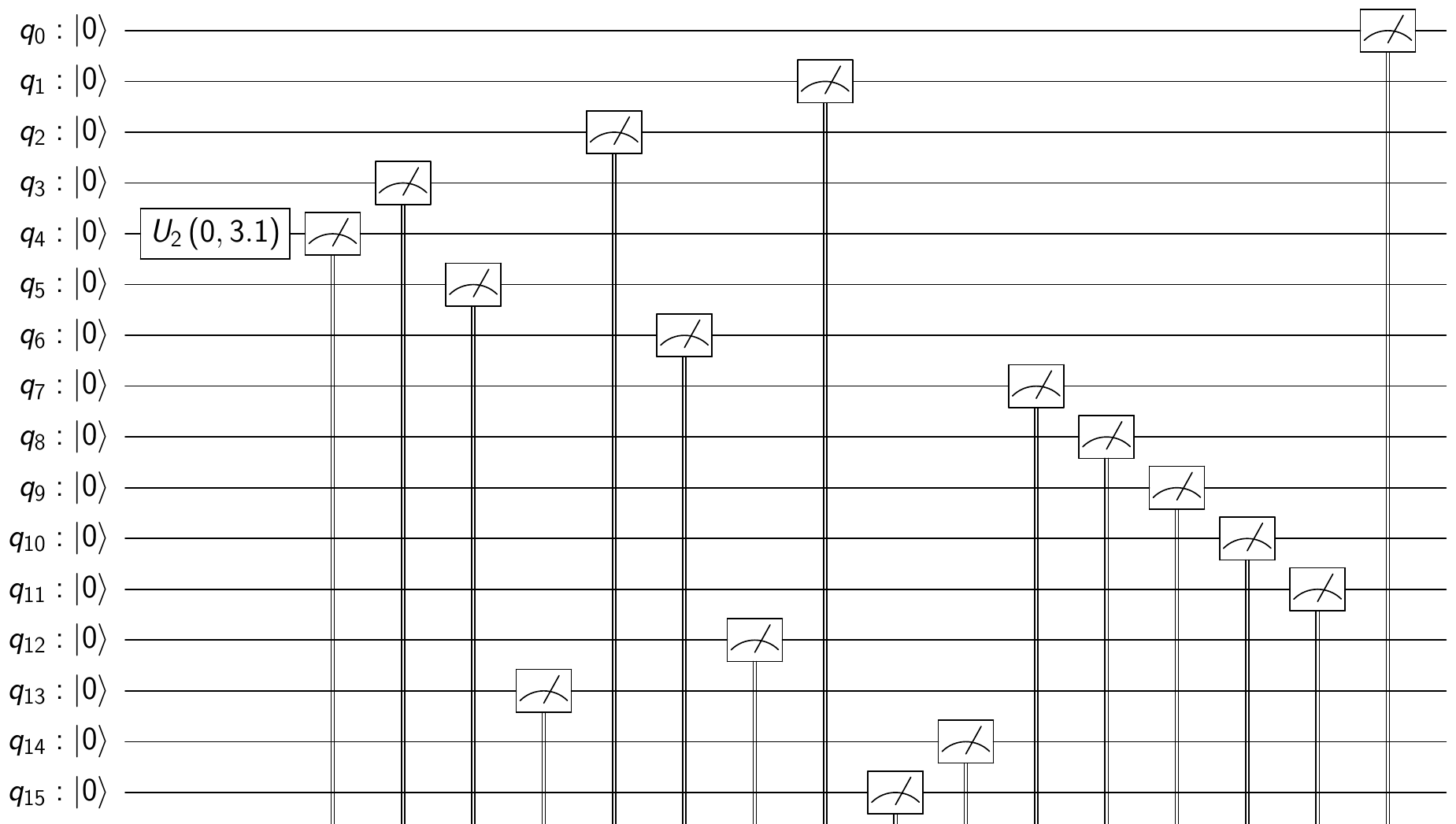}\\
$a=10..10$\\ 
\includegraphics[width=7.5cm]{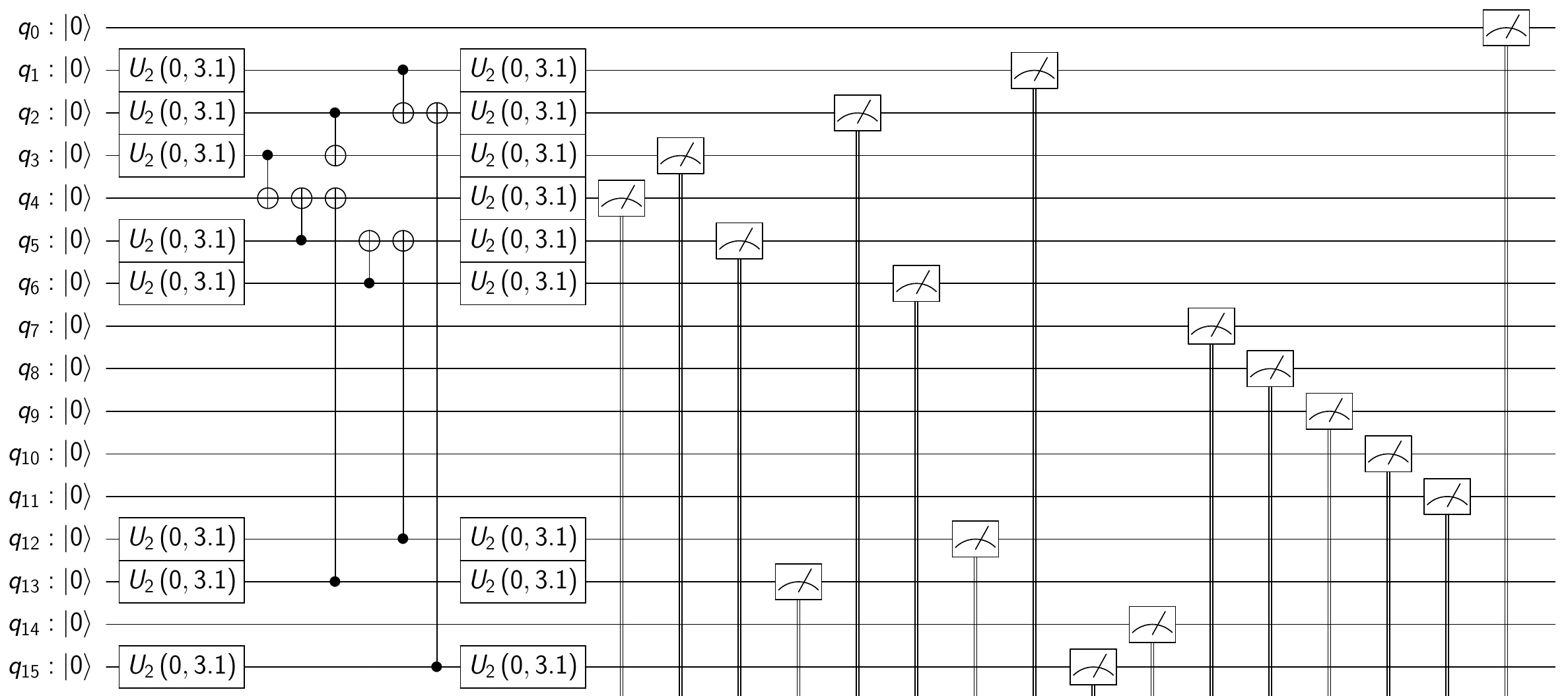}\\
$a=11..11$\\ 
\includegraphics[width=12cm]{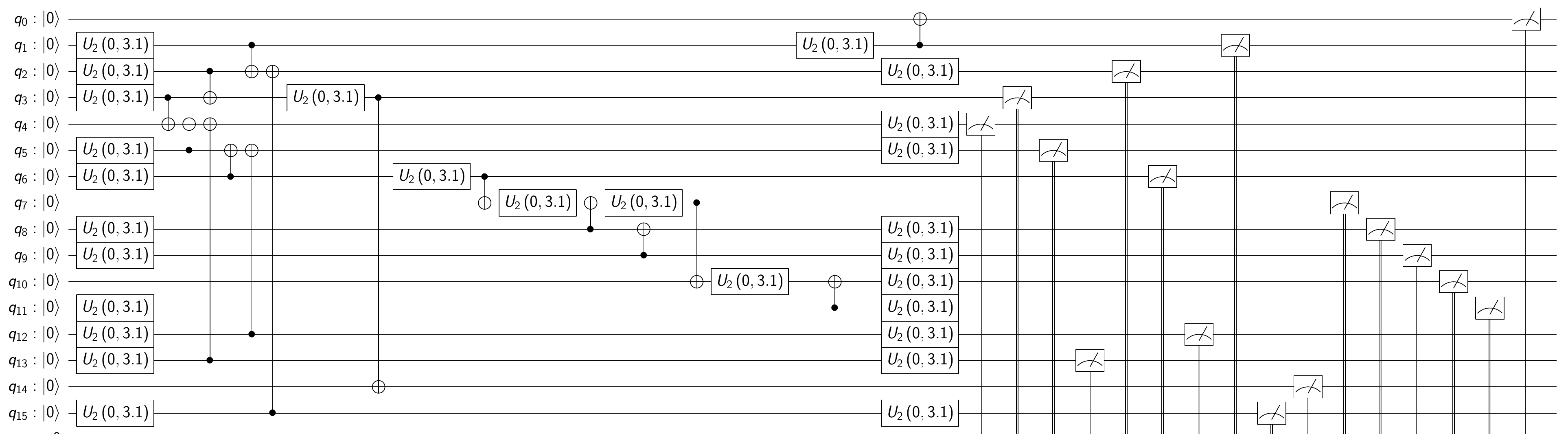}
\caption{Parity learning circuits with quantum oracle encoding $a = 00..00$, $a = 10..10$ and $a = 11..11$ respectively, $n=15$.}
\label{fig:parityCircuits}
\end{figure}

Using QISKit and the compiling strategy described in Section \ref{sec:proposed-strategy}, we implemented three specific circuits on IBM QX5, namely those where the quantum parity oracle encodes $a = 00..00$, $a = 10..10$ and $a = 11..11$. In Figure \ref{fig:parityCircuits}, the circuits corresponding to the case of $n=15$ are illustrated.

In the source code we have released, \textsf{parity.py} can be used to run the experiments on both QX4 and QX5, by setting the number of qubits $n$ and shots $N$.

\subsection{Results}

We have measured the error probability versus number of queries $N$ characterizing the parity learning circuits we implemented on IBM QX5, considering $n=2,8,15$.  

\begin{figure}[!ht]
\centering
\includegraphics[width=8cm]{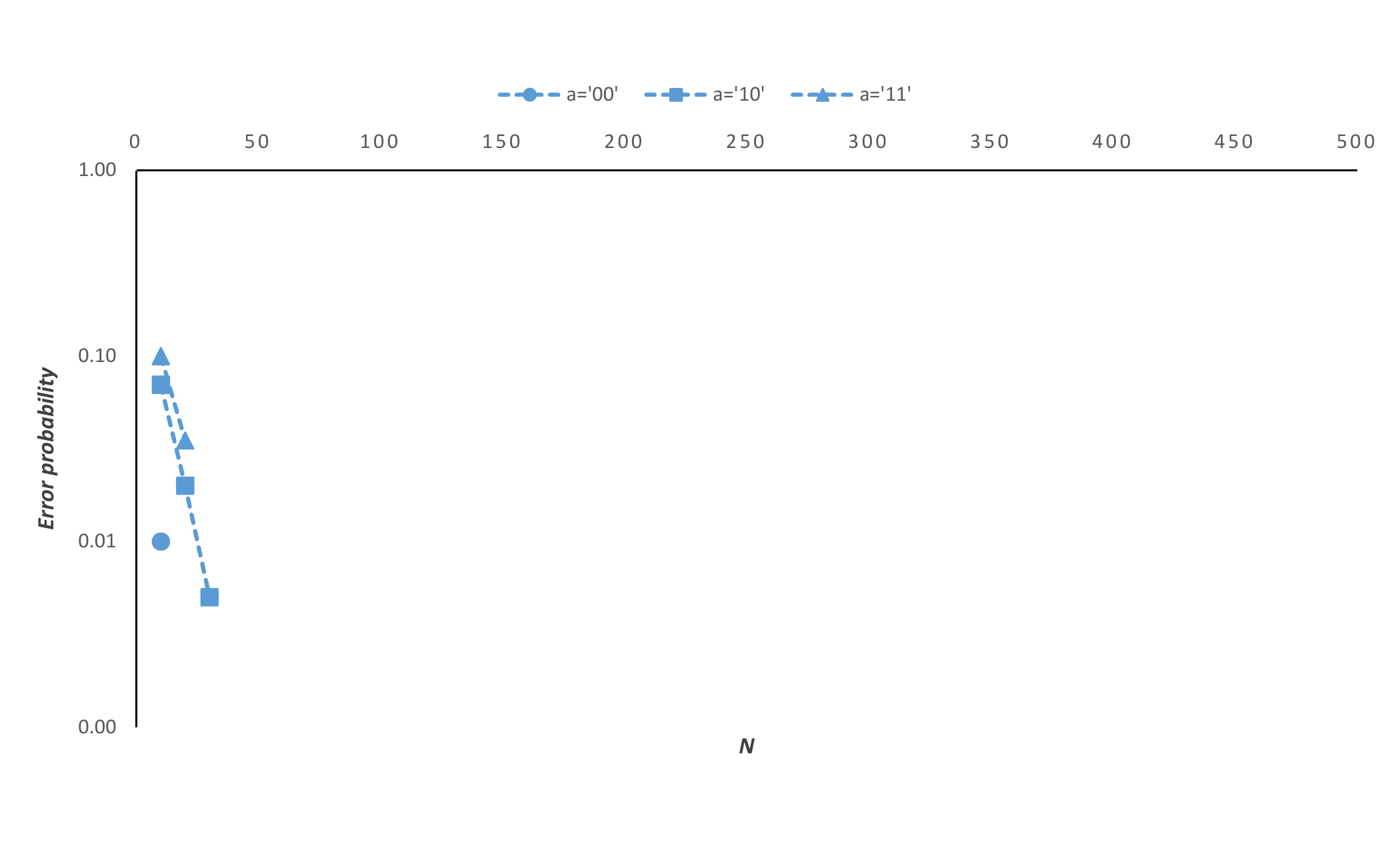}
\includegraphics[width=8cm]{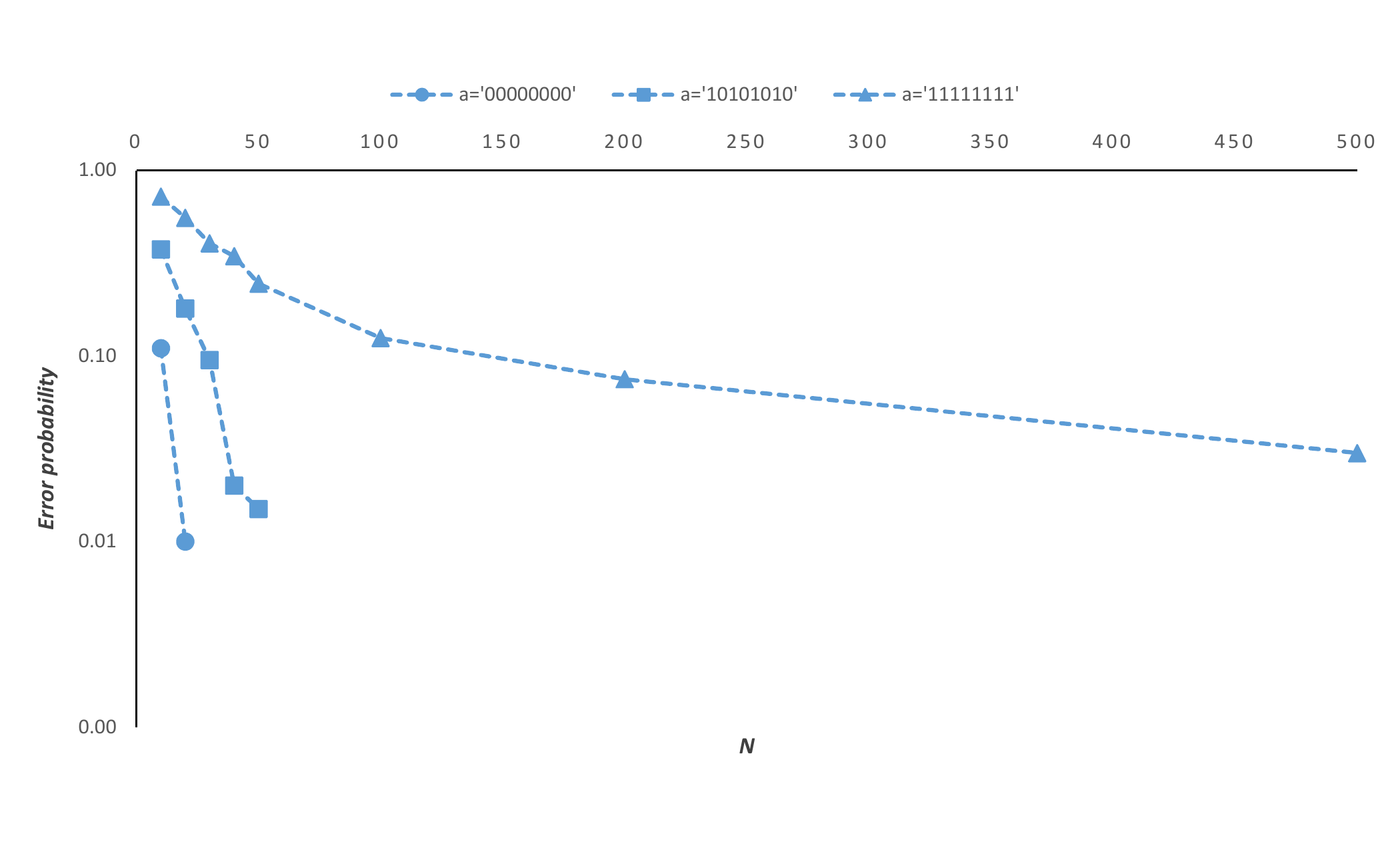}
\includegraphics[width=8cm]{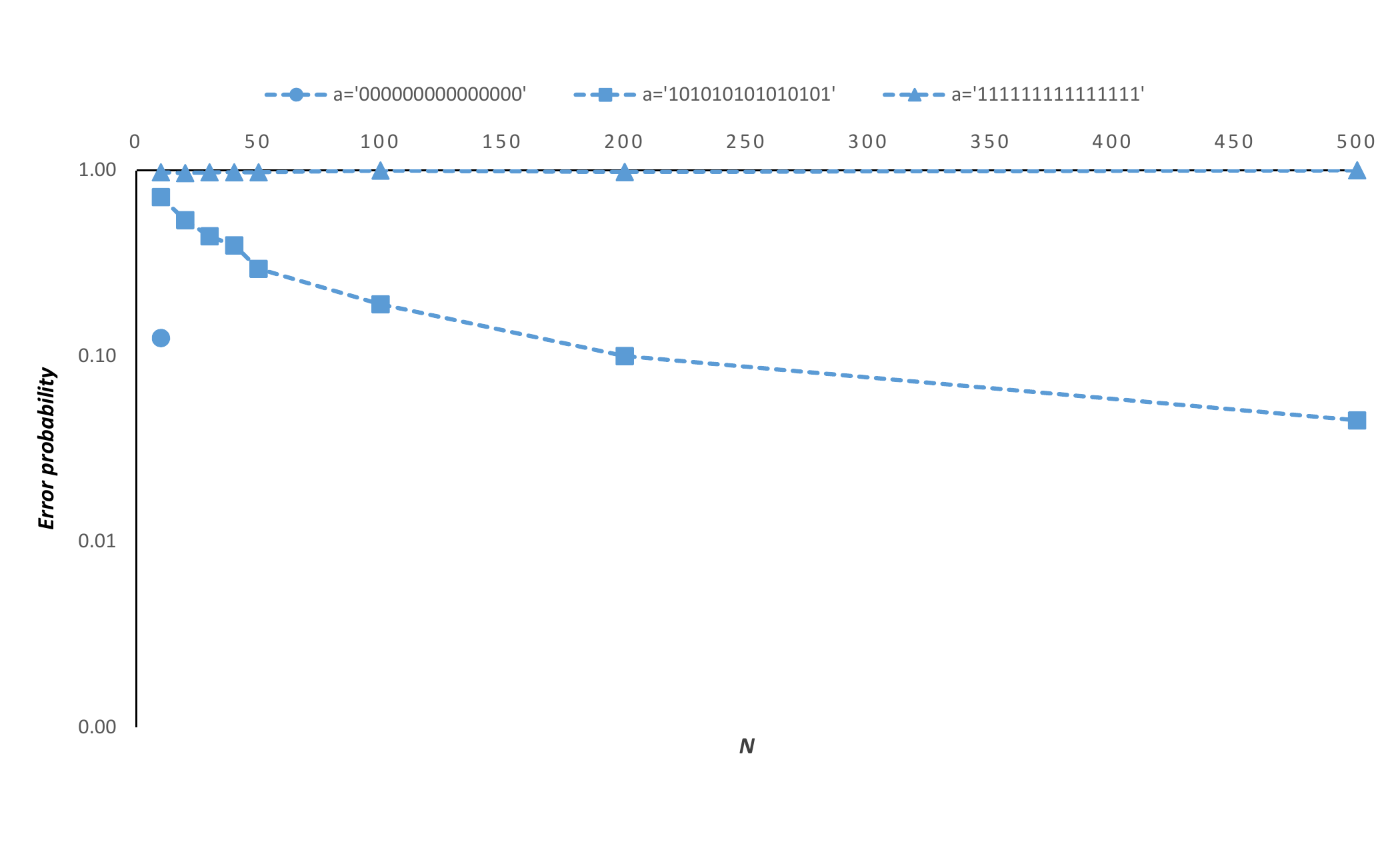}
\caption{Error probability versus number of queries $N$ characterizing the quantum learning circuits with the $n=2,8,15$ qubit quantum parity oracle encoding $a = 00..00$, $a = 10..10$ or $a = 11..11$, on IBM QX5.}
\label{fig:errorProb}
\end{figure}
   
For each $N$ value, the experiment has been repeated $M=200$ times. 
The error probability is the number of successes divided by $M$.
Among the $N$ queries, some give result $1$, others give result $0$.
We postselect on result $1$, and we perform bit-wise majority vote. If the resulting string equals the encoded $a$ string, we count it as a success. We repeat $M$ times and finally we get $p_{err}$ as number of successes versus $M$.

The results are shown in Figure \ref{fig:errorProb}. As expected, $p_{err}$ exponentially decreases with the number of queries $N$, but increases with the number of qubits $n$ and the number of gates. It is worth noting that the 15-qubit quantum oracle encoding $a = 11..11$ is characterized by a nonoptimal choice of one CNOT gate (between qubits $q_6$ and $q_{11}$) in favor of low depth. This is an important lesson for future work.

\section{Concluding remarks}
\label{sec:conclusion}

In this paper, we have illustrated our strategy for compiling low-depth quantum circuits that generate GHZ states. With the same strategy, we have also implemented parity learning circuits. The experimental evaluation of our compiler shows that the produced GHZ circuits are better, in terms of depth and classical fidelity, than those built by the QISKit compiler.

Regarding future work, we will continue in the direction of contributing to the efforts of the QISKit developers community. 
First, we plan to improve our compiling strategy in order to take into account not only the coupling map but also the physical properties of the available qubits and gates, in order to further improve the quality of the resulting quantum circuits. Second, we would like to generalize our compiling strategy to other multi-body entangled states \cite{Motzoi2017} which demand for effective and efficient IBM Q mapping. At a later stage, we will focus on general-purpose quantum compiling, a problem that is still largely open \cite{Venturelli2018,Oddi2018}.

\section*{Acknowledgements}
We acknowledge use of the IBM Q Experience for this work. The views expressed are those of the authors and do not reflect the official policy or position of IBM or the IBM Q Experience team.

We thank Ali Javadi Abhari for his helpful answers to our questions regarding the QISKit compiler.

We thank Diego Rist\`e for his helpful answers to our questions regarding the experimental measurement of error probability in the parity learning demonstration. 

This work has been supported by the University of Parma Research Fund - FIL 2016 - Project ``NEXTALGO: Efficient Algorithms for Next-Generation Distributed Systems''.


\end{document}